\documentstyle[preprint,pra,aps]{revtex}
\begin{document}
\draft

\title{
Problem of a quantum particle in a random potential
on a line revisited
\footnote{Preprint ANU-RsphysSE-20994 \qquad \qquad \qquad
Received 20 September 1994
}
}

\author{O.~K.~Vorov
\footnote{Theoretical Department, School of Physics,
University of New South Wales, Sydney,
NSW 2052, Australia}
and A.~V.~Vagov
\footnote{
Theoretical Physics Department, Research School of Physical Sciences and
Engineering, The Australian National
University, Canberra, ACT 0200, Australia}
}
\date{20 September 1994}
\maketitle
\begin{abstract}
The density of states for a particle moving in a random
potential with a Gaussian correlator is calculated exactly using
the functional integral technique. It is achieved by expressing the functional
degrees of freedom in terms of  the spectral variables and the
parameters of isospectral transformations of the potential. These
transformations are given explicitly by the flows of the
Korteweg-de Vries
hierarchy which deform the potential leaving all its spectral properties
invariant. Making use of conservation laws reduces the initial
Feynman integral to a combination of quadratures which can be readily
calculated.
Different formulations of the problem are analyzed.
\end{abstract}
\pacs{PACS: 71.20.-b,71.20.Ad,71.25.Mg,71.55.Jv}

The problem of a quantum particle moving in a random potential has been
of interest for a long time. It has numerous applications to solid
state physics \cite{GENERAL},
particularly to the electron structure of heavily doped semiconductors.
Various modifications of the problem of the calculation of the Green's
function and Density of States (DOS) averaged over the ensemble of
random potentials arise (Thouless and Edwards \cite{THOULESS})
in connection with the Anderson localization
\cite{ANDERSON}, in the physics of amorphous materials, particularly glasses
\cite{PARISI}, in considerations of the conductivity in wires and mesoscopic
objects \cite{WEIDENMULLER}, in the phenomena of broadening of resonance
lines in solids \cite{LIFSHITZ,STONEHAM}, in a number of problems
connected to the Quantum Hall effect \cite{SA-YAKANIT} and many
other problems related to disorder\cite{WEGNER,GOLDSHTEIN,ALTSHULER}.

In particular, the one-dimensional problem with a random potential having
a white-noise correlator has been considered in a large number of
works. The case when the potential is arbitrary but varies very slowly
compared with the wave function of an electron was considered by Kane
\cite{KANE}. To calculate the DOS for the reverse situation with
randomly distributed $\delta$-functional wells, Frish and Lloyd employed
stochastic equations \cite{FRISCH}. In the subsequent studies of the more
general case of an arbitrary uncorrelated potential the methods of
``minima counting'' (Halperin and
Lax \cite{HALPERIN}) and ``zeros counting" were used
(Lifshits \cite{LIFSHITZ}, Halperin \cite{HALPERIN2}). Zittars and Langer
formulated it
in the Feynmann path integral language \cite{ZITTARS}. That approach was
further developed by Parisi who used the replica trick \cite{PARISI} and
alternatively by Efetov by means of auxiliary Grassmannian fields
\cite{EFETOV}.
Berezinski rigorously considered the problem
by a diagram summation technique \cite{Berezinski}.
Luttinger and Waxler considered special
potential configurations ($\delta$-functional scatterers) by the
functional integral technique \cite{LUTTINGER}.

However, even for the 1D problem with a white-noise correlator the
average density of states $\eta(E)$ has not been found exactly due to
technical difficulties specific for the methods used. Most of the
results obtained so far give only the ``tail'' behavior of the
density of states, i.e., the DOS asymptotics at large negative
energies $E$ which is found to be $\sim exp(-|E|^{3/2})$.

A nonperturbative approach to the problem presented here uses essentially
the ideas of Inverse Scattering Method (ISCM) for  solving nonlinear
evolution equations proposed in the celebrated works by Gardner, Greene,
Kruskal and Miura \cite{GARDNER}, and Lax \cite{LAX} and developed
in Refs.\cite{SHABAT,ZAKHAROV,NEWELL,ESTABROOK,FLASCHKA,TAKHTAJAN}.
Methodically, what
is done in this work is a kind of inversion of this
'diamond of mathematical physics'. Here, the Schr\"odinger equation is
to be solved and we use the hierarchy of Korteweg-de Vries (KdV) equations
to find a parametrization of the random potential.

DOS as
a function of energy $E$ averaged over the ensemble of random potentials
with a Gaussian white noise correlator can be found
from the functional integral (FI):
\begin{eqnarray}
\label{FI}
\langle \eta (E) \rangle =
\frac{1}{\pi} \lim_{\delta \rightarrow 0}
Im[ Tr \langle G_{E-i\delta} \rangle ] =
\qquad \qquad
\nonumber\\
\frac{1}{\pi Z} \lim_{\delta \rightarrow 0} \int D[U(x)] e^{-a^3 \int
U(x)^2dx}
Im[
Tr G_{E-i\delta} \{U(x)\} ] ,
\end{eqnarray}
where $G_{E}\{ U(x)\}$ is the Green's function of a particle in the
potential $U(x)$, $D[U(x)]$ is a functional integration measure,
$\langle...\rangle$ means an ensemble average. The partition function $Z = \int
D[U(x)] e^{-a^3 \int U(x)^2dx}$ is the normalization constant, the trace
means the integral $Tr [G] = \int_{-\infty}^{\infty} dx G(x,x'=x)$ and the
constant $a$ is the ensemble correlation length of the dimensionality
$[x]=[E^{-1/2}]$.
We use the units in which $\hbar^2 /m =1$.
Due to the Gaussian weight only
quadratically normalizable configurations enter the path integral (\ref{FI}):
\begin{equation}
\label{COND}
\int_{-\infty}^{+\infty} U^{2}(x)dx < \infty .
\end{equation}
Since the functions $U(x)$ model
the real impurity potentials,
% (impurity potential of a real system?),
which are smooth, it is
assumed that $U(x)$ can be differentiated an appropriate amount of
times.

The Green's function $G$ in (\ref{FI}) corresponds to the
eigenvalue problem for the Schr\"odinger operator ${\hat L}$
\begin{eqnarray}
\label{SC}
[E - \hat L] G_{E}(x,x') = \delta(x-x'), \quad
\hat L \psi_{k}(x) = \varepsilon_{k} \psi_{k}(x),
\nonumber\\
%\qquad
\hat L = - \frac{1}{2} \frac{d^{2}}{dx^{2}} + U(x)
\qquad \qquad \qquad \qquad
\end{eqnarray}
with the boundary conditions specified below. To average the DOS
one has to solve these equations for every particular potential
configuration $U(x)$ entering the path integral and that appears to be
the main difficulty in a straightforward approach. Let us note that the
present problem (\ref{FI})
is similar to the one considered in the Random Matrix Theory
(RMT) \cite{PORTER}.

The boundary conditions for Eq.(\ref{SC}) are dictated by
requirement (\ref{COND}) and are typical for the scattering problem
\footnote{Actually, the
stricter condition $\int |U(x)|(1+|x|)dx < \infty$ is required for
rigorous validity of the following consideration. It means the potential
vanishes rapidly at infinity.}.
The particle is moving on a line in a
square normalizable potential vanishing at infinity. Motion at
$E>0$ is unbounded, and the spectrum remains continuous as for a free
particle.  Therefore the density of states at $E>0$ remains unaffected
by such a potential  and equals $const \times E^{-1/2}$. Thus, we are
interested in DOS for negative energies where all states are bounded and
have discrete eigenvalues. So the boundary conditions for Eq.(\ref{SC})
are assumed to be $\psi (\pm \infty )=0$.

Our method employs an accurate change of variables in the path integral
(\ref{FI}). The idea is to use the spectral data of a potential as a set
of new variables instead of the potential itself. The spectral data alone
cannot describe a potential configuration uniquely,  because a lot of
other configurations entering the integrand (\ref{FI}) yield the same
spectral properties. To account for all the potentials from the same
isospectral family one needs to have an additional set of variables
which describe all the spectrally
equivalent configurations.
The generators of the group of the isospectral transformations of the
Schr\"odinger operator ${\hat L}$
are the Lax operators ${\hat A}_{2j+1}$
obeying the commutation relations \cite{LAX}
\begin{equation}
\label{LAX}
[d/d \tau_{2j+1} + {\hat A}_{2j+1},{\hat L}] = 0,
\qquad [{\hat A}_{2j+1},{\hat A}_{2j'+1}]=0,
\end{equation}
where $j$ is a positive integer and
the ``times" $\tau_{2j+1}$ play the role of parameters and count the
members of an isospectral family of potentials. The relations
(\ref{LAX}) enforce the set nonlinear equations for $U(x)$
(KdV hierarchy)
\begin{equation}
\label{HIER}
U_{\tau_{1}}=U_{x}, \qquad U_{\tau_{3}}=-U_{xxx}+12UU_{x}, \quad...
\end{equation}
(the notation $U_{a} \equiv \partial U/ \partial a$ is used).
The role of parameters $\tau_{1}$,$\tau_{3}$,$...$
is analogous to that of the angles of unitary
transformations in the language of the RMT \cite{PORTER}.

The essence of the ISCM theory (see, e.g., \cite{SHABAT}-\cite{NEWELL})
which is important in the present context
can be reformulated in the following rigorous statements:

(i) Each square normalizable potential configuration, provided that
condition (\ref{COND}) is fulfilled (see footnote 1)
\footnote{ This requirement is
important for the establishment of analytical properties of solutions of
(2) that used in the ISCM \cite{NEWELL}.} can be characterized by an
integer number of bound states $N \geq 0$, with the set of
energies $\varepsilon_{i}= \kappa_{i}^{2}, 1 \leq i \leq N$,

(ii) Each solution of the KdV hierarchy consists of a soliton component
$U_{s}$ controlled by the finite number of discrete eigenvalues of the
problem (\ref{SC}) and a radiation component $U_{r}$ described by a
continuous set of variables related to the reflection coefficient $R$.
The soliton and radiation degrees of freedom are independent.

(iii) The possibility for a configuration to evolve according to the KdV flows
(\ref{HIER}) gives rise to a corresponding volume of isospectral
transformations associated with it. The total weight of the configurations
with definite set $\{\varepsilon_{i},N\}$ in the ensemble is
proportional to this volume.

Each of the potential configurations
that enter our functional integral
viewed as a solution of the KdV hierarchy (\ref{HIER}) and it can be
obtained from the trivial configuration $U(x) \equiv 0$ by means of the
B\"acklund transformations
%\cite{ESTABROOK}
or by Hirota's direct
method \cite{ESTABROOK}.

In view of the above classification (i)-(iii),
the two following formulations of the
problem of calculation of the average DOS are natural:

{\bf A.} The total number of bound states allowed for a given ensemble
is fixed $N \{ U(x) \} = N =const$ for all particular realizations of
the random potential. In that case, one should calculate the total level
density $\eta (E)$.

{\bf B.} The total number of discrete levels is not fixed by physical
conditions, and calculation of the ``reduced'' DOS, i.e., the density
per a length unit, $\eta' (E) =d \eta (E)/d x$, is reasonable.
In this case, one should keep the total ``concentration'' of
bound states on the line $c$ fixed.

We shall consider both cases, implying that a
particular choice should follow from the physical requirements for a
given system under consideration.

We evaluate the functional integral (1) using the exact parametrization of the
potential viewed as a solution of the KdV hierarchy (5), that is given
by the complete set of new variables \cite{ZAKHAROV,FLASCHKA,NEWELL}
\begin{equation}
\label{data}
U(x) =
U(x,\underbrace  { \varepsilon_{i}, \{ {\cal P}(k) \}  }_{\it actions},
\underbrace{ \theta_{i}, \{ {\cal Q}(k) \} }_{\it angles} )
\end{equation}
where $\varepsilon_{j}<0$ are the energies of the ``solitons'' (bound states),
$\theta_{j}$ are soliton phases related to the soliton positions
$X_{j}$ as
$\theta_{j}$$=$$|$$2$$\varepsilon_{j}|^{1/2}$$X_{j}$$=$$\sum_{m=0}^{\infty}$$
|$$2$$\varepsilon_{j}$$|$$^{m+1/2}$$\tau_{2m+1}$,
and the continuous set ${\cal Q}(k)$ defined at positive energies
$\varepsilon=2k^{2}$
describes the isospectral transformations of the radiation part.
The radiational {\it actions} set ${\cal P}$$(k)$ is related to
the reflection coefficient:
${\cal P}$$(k)$$=$$-$$\frac{2k}{\pi}$$log$$($$1$$-$$|$$R$$($$k$$)$$|^{2})$.
As the KdV hierarchy is a fully integrable Hamiltonian system
\cite{ZAKHAROV}, Eq.(\ref{data}) is the canonical transformation from
the potential variables $U(x)$ with given number $N$ of bound states to the
action-angle type variables \cite{FLASCHKA}.  The canonical Hamiltonian
structure of the KdV hierarchy is governed by the standard Poisson brackets
for the new variables $\{ \varepsilon_{j},\theta_{j'}\}=\delta_{jj'}, \quad
\{{\cal P}(k),{\cal Q}(k')\}=\delta(k-k')$ with all other brackets zero
\cite{TAKHTAJAN}.
The Jacobean of the canonical transformation
$U(x)\rightarrow \{\varepsilon_{j},\theta_{j},{\cal P}(k),{\cal Q}_{j}\}$
is unity \cite{MECHANICS} and the
measure is just a product of differentials of the new variables. This
transformation is correct because only smooth configurations are to be
counted in the Feynmann integral and the problem with discontinuous
paths typical in the field theories is irrelevant in the
present context \cite{BULLOUGH}.
Thus we can express the functional measure in
(\ref{FI}) in terms the new variables as follows
\begin{eqnarray}
\label{MEASURE}
\int D [U(x)] \Phi \{ U(x) \}=
\int \prod_{x}dU(x) \Phi \{U(x) \}=
\qquad \qquad
%\nonumber
\\
\int\prod_{j=1}^{\infty}d\varepsilon_{j}d\theta_{j}
D[{\cal P}(k)]D[{\cal Q}(k)]
%\quad
\Phi\{U(x,\varepsilon_{j},\theta_{j},\{{\cal P},{\cal Q}\})\}, \qquad \nonumber
\end{eqnarray}
where $\Phi $ is the integrand of Eq.(\ref{FI}).
Let us stress that while we integrate over parameters of the
evolution of the KdV equations we count all but the {\it static} potential
configurations, and no real time evolution is involved in the problem at
all.

The next advantage of the soliton language is the possibility to
rewrite the Gaussian weight in Eq. (\ref{FI})
compactly using the second conservation law for the KdV hierarchy
\cite{NEWELL}
\begin{equation}
\label{integral}
\int \limits_{-\infty}^{\infty}U(x)^2dx=
\sum_i^{N}\frac{4}{3}|2\varepsilon_{i}|^{\frac{3}{2}}
+
\int \limits_{0}^{\infty}\frac{2k^{2}}{\pi}log(1-|R(k)|^2)dk.
\end{equation}

First we consider the case {\bf A} when the total number $N$ of bound
states is fixed. The expression (\ref{FI})
with the account for (\ref{MEASURE}),(\ref{integral})
read as follows
\begin{mathletters}
\label{ZN}
\begin{eqnarray}
\langle \eta (E) \rangle
%&&
= \frac{I_N}{Z_N} =
\frac{1}{Z_N N!}\int D[{\cal P}(k)]D[{\cal Q}(k)] \int \prod_{i}
d\varepsilon_{i}d\theta_{i} \nonumber\\
%&&
\times
\exp [-4a^{3}\sum_{i}\frac{\varepsilon_{i}^{\frac{3}{2}}}{3}
-
a^{3}\int_{0}^{\infty}kdk {\cal P}(k))]
\sum_{i=1}^{N}\delta(E-\varepsilon_{i}), \qquad
\end{eqnarray}
with
\begin{eqnarray}
Z_N =\frac{1}{N!} \int D[{\cal P}(k)]D[{\cal Q}(k)] \int \prod_{i}
d\varepsilon_{i}d\theta_{i} \qquad
\nonumber\\
\times
\exp[-\frac{4}{3}a^{3}\sum_{i}|2\varepsilon_{i}|^{3/2}
-a^{3}
\int_{0}^{\infty} k {\cal P}(k)dk].
\end{eqnarray}
\end{mathletters}
The $(N!)^{-1}$ factor appears due to symmetry reasons.
The integration over
the ``radiation'' variables $\{ {\cal P}(k), {\cal Q}_(k) \}$
corresponds to the continuous spectrum of ISCM and is still continual.
It is important that when expressed through the conserving quantities
of the KdV equations,
the integrands in (\ref{ZN}) depend only on the {\it action} variables $\{
\varepsilon_{i}, {\cal P}(k) \}$.

The functional integration over ${\cal P}(k)$ and ${\cal Q}(k)$
factorizes giving the same functional constant
$\int$$D[{\cal P}(k)]$$D[{\cal Q}(k)]$
$\exp[-a^{3} \int_{0}^{\infty}k{\cal P}(k)dk]$
for both the numerator $I_N$ and the denominator $Z_N$ in the Eq.(\ref{ZN}).
It is the manifestation of the fact discussed above that the
radiation component of the potential $U_{r}$$($$x$$)$ does not yield bound
states and does not affect the ``soliton energies'' $\varepsilon_{i}$.
Hence it does not influence the DOS for $E$$<$$0$.
To define the divergent integrations over the soliton phases $\theta_{i}$
we introduce the length of the system to play the role of a cutting parameter.
In order to do so
we rewrite the
remaining integrals in the Eq.(\ref{ZN}) in terms of the soliton
coordinates
$X_i=\theta_i/|2\varepsilon_i|^{1/2}$ \cite{NEWELL}. The Jacobean for this
transformation is $\prod_{i=1}^{N} \sqrt{2|\varepsilon_{i}|}$ and we obtain
from (\ref{ZN})
\begin{equation}
%\begin{eqnarray}
\label{SOL}
\langle\eta(E)\rangle=
\frac{\prod_{i}^{N} [\int\limits_{0}^{\infty}d\varepsilon_{i}
|2\varepsilon_{i}|^{\frac{1}{2}}\int\limits_{-\infty}^{\infty}dX_{i}]
exp[-\sum_{i=1}^{N}\frac{4|2\varepsilon_{i}|^{3/2}a^{3}}{3}]
\sum_{i=1}^{N}\delta(E-\varepsilon_{i})}
{\prod_{i}^{N}[\int\limits_{0}^{\infty}d\varepsilon_{i}
|2\varepsilon_{i}|^{\frac{1}{2}}\int\limits_{-\infty}^{\infty}dX_{i}]
exp[-\frac{4}{3}a^{3} \sum_{i=1}^{N}|2\varepsilon_{i}|^{3/2}]}
%\begin{eqnarray}
\end{equation}

It is interesting that similar expressions are normally obtained in the
Gaussian ensemble of random matrices (except for the change
$\varepsilon_i^{2}\rightarrow \varepsilon_i^{3/2}$) \cite{PORTER}. However,
the
level repulsion is absent here contrary to the RMT, where
it usually appears in the form of the factor $\prod_{i\neq
j}|\varepsilon_i-\varepsilon_j|$ \cite{PORTER}.
This property is specific for the
the infinite length line, where one can have a potential
configuration with two bound states with an arbitrary small difference
between corresponding eigenvalues. Now all the integrals over the
isospectral variables $X_i$ cancel, and Eq. (\ref{SOL}) is
reduced to the combination of quadratures which gives the final result
for the local DOS
\begin{equation}
\label{A}
Case \quad {\bf A}:
\langle \eta (E) \rangle =
4 a^3 N \sqrt{|2E|} e^{ -\frac{4}{3}a^{3}
|2E|^{3/2} }.
\end{equation}

We turn now to the case {\bf B}, where the number of bound states is not
fixed from the very beginning.
Due to the above classification every potential configuration
belongs to one of the classes  which differ by the number of the solitons
$N$. And instead of (\ref{ZN}) we can write the
following expression for the averaged DOS
\begin{equation}
\label{STAT}
\langle \eta (E) \rangle =
\frac{D_{0}I_{0}+D_{1}I_{1}+ ... + D_{N}I_{N} + ...}
{D_{0}Z_{0}+D_{1}Z_{1}+ ... + D_{N}Z_{N} + ...}
\end{equation}
with the integrals $I_{N}$,$Z_{N}$ given by Eq.(\ref{ZN})
\footnote{Note the formal similarity between present
classification and those arising in the nonperturbative field theories
in discussions of the multi-instanton configurations
\cite{POLYAKOV}.}.
The dimensional functional constants $D_{N}$ correspond to the
relative probabilities for the N-soliton configurations to occur.
To evaluate the expression (\ref{STAT}) we recall that creation
of a new $N+1$'th soliton in addition to $N$ existing ones
gives rise to an additional degree of freedom. It is related to ``translation''
of this extra potential well along the $X$ axis. It enlarges the weight
of $N+1$-soliton potential configurations by a formally divergent
factor $\int dX$.
Thus the configurations containing the number
of solitons less than a given integer $N$ have a zero measure
in the ensemble compared to $N$-soliton configurations,
and Eq.(\ref{STAT}) reduces to
\begin{equation}
\label{STAT1}
\langle \eta (E) \rangle = \lim_{N \rightarrow \infty}
\frac{D_{N} I_{N} } {D_{N} Z_{N} } =
\lim_{N \rightarrow \infty}
\frac{ I_{N} } {Z_{N} }.
\end{equation}
The fact that only the terms with maximum $N$ survive in Eq.(\ref{STAT})
can be proved explicitly by considering a discretized version of FI.
The probabilities for $N$-soliton potentials to appear
are evaluated with the aid of the expression
for the number of bound states (solitons)
$N = \frac{1}{\pi} \int_{U<0} \sqrt{|U(x)|}dx$ \cite{KARPMAN}
valid for large $N$
which we need in Eqs.(\ref{STAT},\ref{STAT1}).

The total average DOS is related to the reduced DOS
(i.e., per unit length) $\eta'(E)$
by the obvious relation $\eta(E) = \int dx \eta'(E)$. The reduced DOS is
independent of $x$ due to translational invariance.
After the simple transformations we obtain the result
\begin{equation}
\label{B}
Case \quad {\bf B}: \quad
\langle \eta'(E)\rangle =
4a^{3}c
\sqrt{2|E|}e^{-a^{3}\frac{4}{3}
|2E|^{3/2}}
\end{equation}
The total concentration of the bound states on the line, $c$,
mentioned above,
appears here as a natural limit
$N /\int dx \rightarrow c$ at very large $N$.

Note that both the expressions (\ref{A})
and (\ref{B}) for the DOS correspond to the ``local'' density which is
connected to the so called ``cumulative'' density of states ${\cal
N}(E)$ by the relation $\eta(E) = d {\cal N}(E)/dE$
\cite{HALPERIN,ZITTARS}. The value ${\cal N}(E)$ is the number
of levels with energy less than a given value $E$.
Expressions (\ref{A}) and (\ref{B}) are similar in form to those
obtained elsewhere \cite{FRISCH,HALPERIN} for the asymptotic of the DOS
in the large $|E|$ limit (band tails). However in the present approach
no approximation has been used, and therefore the result is valid for
all energies.

To summarize, in this Letter we presented
a nonperturbative method to calculate
the DOS in a random potential. Using the exact
classifications of the potential configurations in the
functional integral in terms of the inverse scattering method,
we obtain an analytical result for the DOS random
potential with a white-noise correlator. No restriction by
particular energy regions are needed within this approach.

This method can be applied, though
with some modifications, to a calculation of the averaged Green's
function $\langle G(x,x') \rangle$ (two-point correlator).
This will be presented in a forthcoming paper.

Concluding,
we note that the present approach
is based on a deep connection between the 1D
Schr\"odinger equation and the exactly solvable nonlinear KdV hierarchy
and hence can be directly applied to 1D problems only. But the method
of using isospectral transformations to fulfill the averaging over
the field ensemble looks quite fruitful and we believe that it can be
done in a number of related problems in more than one dimensions in
particular for the calculation of the DOS in a 2D random potential in the
presence of magnetic field.
It can also be used in
random matrix theory. Work in these directions is now in progress.

{\bf Comment added after submission:}
After our paper has been submitted to this database
we were made known that a similar
approach to the problem was discussed earlier by
B.N.Shalaev in Fiz. Tverd. Tela (Leningrad) {\bf 32} 3586 (December 1990),
[Sov. Phys. Solid State {\bf 32 (12)} 2079 (1990)] (Ref. \cite{SHALAEV}).
However, important aspects of considerations and the results of the
two works are different.
We are grateful to R.H. McKenzie for making us known about
the work by B.N.Shalaev.

\noindent

\begin{thebibliography}{200}
\bibitem{GENERAL}
V.L.Bonch-Bruevich,
{\it The Electronic Theory of Heavily
Doped Semiconductors}, N.Y., American Elsevier, 1966;
J.M.Ziman, {\it Models of Disorder}, Cmbr.Uni.,
Cambridge, 1990;
P.V.Mieghem, Rev. Mod. Phys. {\bf 64}, 755 (1992).

\bibitem{THOULESS}
D.J.Thouless, Phys.Rev.Lett. {\bf 39}, 1167 (1977);
J. Phys. {\bf C5}, 77 (1972), {\it ibid}. {\bf C6}, (1973);
J.T.Edwards and D.T.Thouless, J.Phys. {\bf C5}, 807 (1972).
\bibitem{ANDERSON}P.W.Anderson, Phys.Rev. {\bf 109}, 1492 (1958).

\bibitem{PARISI}
G.Parisi, {\it
Field Theory, Disorder and Simulations}, Wrld. Sci., Singapore,
1992, and references therein.

\bibitem{WEIDENMULLER}S.Iida, H.A.Weidenm\"uller and J.A.Zuk,
Phys.Rev.Lett. {\bf 64}, 583 (1990) and Ann.Phys. (NY) {\bf 200}, 219
(1990).

\bibitem{LIFSHITZ}I.M.Lifshitz, Sov. Physics Uspekhi, {\bf 7}, 549
(1965), [Sov.Usp.Fiz.Nauk {\bf 83}, 617 (1964)].

\bibitem{STONEHAM}A.M.Stoneham, Rev.Mod.Phys. {\bf 41}, 82 (1969).
\bibitem{SA-YAKANIT}V.Sa-yakanit and H.R.Glyde, in Proc. of the {\it
Third International Conference on Path integral from $meV$ to $MeV$},
Wrld.Sci., Singapore, 1989, p. 163; K.Esfarajani, H.R.Glyde, and
V.Sa-yakanit, ibid. p.176, see also R.E.Prange, in {\it Quantum Hall
Effect}, R.E.Prange and S.M.Girvin eds, Springer-Verlag, 1987.
\bibitem{WEGNER} F.J.Wegner, Phys.Rev. {\bf B19}, 783 (1979);
K.Ziegler, Phys.Lett. {\bf 92A}, 339 (1982).

\bibitem{GOLDSHTEIN}Ya.Gol'dshtein, S.A.Molchanov and L.A.Pastur,
Func.Anal. \& Appl.{\bf 11}, N1, 1 (1977).

\bibitem{ALTSHULER} B.D.Simons and B.L.Altshuler, Phys.Rev.Lett.
{\bf 70}, 4063 (1993);
A.G.Aronov and P.W\"olfe, {\it ibid}. {\bf 72}, 2239 (1994).

\bibitem{KANE}E.O.Kane, Phys.Rev. {\bf 131}, 79 (1963).
\bibitem{FRISCH}H.L.Frisch and S.P.Lloyd, Phys.Rev. {\bf 120}, 1175
(1960); B.Halperin, Phys.Rev. {\bf 139}, A104 (1965);
T.P.Eggarter, Phys.Rev. {\bf B5}, 3863 (1972).

\bibitem{HALPERIN}B.I.Halperin and M.Lax, Phys.Rev. {\bf 148}, 722
(1966).

\bibitem{HALPERIN2}B.I.Halperin, Adv.Chem. Phys. {\bf 13}, 123 (1967).

\bibitem{ZITTARS}J.Zittars and. J.S.Langer, Phys.Rev. {\bf 148}, 741
(1966).

\bibitem{EFETOV} K.B.Efetov, Adv.Phys. {\bf 32}, 53 (1983).

\bibitem{Berezinski}V.L.Berezinskii, Zh.Eksp.Teor.Fyz. {\bf 65}, 1251
(1973) [Sov.Phys.JETP {\bf 38}, 620 (1974)]

\bibitem{LUTTINGER}J.M.Luttinger and R.Waxler, Ann.Phys. {\bf 175}, 319
(1986).

\bibitem{GARDNER}C.S.Gardner, J.M.Greene, M.D.Kruskal and K.K.Miura,
Phys. Rev. Lett. {\bf 19}, 1095 (1967).

\bibitem{LAX}P.D.Lax, Comm. Pure Appl. Math. {\bf 21}, 467 (1968).

\bibitem{SHABAT}V.E.Zakharov and A.B. Shabat, Func.Anal. \& Appl. {\bf
8}, 43 (1974).

\bibitem{ZAKHAROV}V.E.Zakharov and L.D.Faddeev, Func.Anal. \& Appl. {\bf
5}, 18 (1971).

\bibitem{NEWELL} A.C.Newell, {\it Solitons in Mathematics and Physics},
Philadelphia, SIAM, 1985.

\bibitem{ESTABROOK}H.D.Wahlquist and F.B.Estabrook, Phys.Rev.Lett. {\bf
31}, 1386 (1975);
R.Hirota, Phys.Rev. lett. {\bf 27}, 1192 (1972).

\bibitem{FLASCHKA}H.Flaschka and A.C.Newell, in {\it Dynamical systems,
Theory and Applications}, J.Moser ed., Lecture Notes in Physics {\bf
38}, Schpringer-Verlag, N.Y., 1975, p.355.

\bibitem{TAKHTAJAN}L.D.Faddeev and L.A.Takhtajan, Lett. Math. Phys. {\bf
10}, 183 (1985).

\bibitem{PORTER} {\it Statistical Theories of Spectra: Fluctuations}.
C.E.Porter ed., Acad. Press, N.Y. \& London, 1965;
M.L.Mehta, {\it Random Matrices}, 2nd ed., Boston, Acad.
Press, 1991, see also
%\bibitem{ZIRNBAUER}
M.R.Zirnbauer, J.J.M.Verbaarschot, and
H.A.Weidenm\"uller, Nucl.Phys. {\bf A411}, 161 (1983).

\bibitem{MECHANICS}
H.Goldstein, {\it Classical Mechanics}, Cambridge, Mass.,
Addison-Wesley, 1953.
\bibitem{BULLOUGH}R.Bullough, D.J.Pilling and J.T.Timonen, in
{\it Dynamical Problems in Soliton Systems}, S.Taneko ed.,
Springer, Berlin 1985,
p.105;
K.Sasaki, {\it ibid}. p.122.

\bibitem{POLYAKOV}A.M.Polyakov, Nucl.Phys.B {\bf 120}, 429 (1977).

\bibitem{KARPMAN} V.I.Karpman and V.P.Sokolov,
Zh.Eksp.Teor.Fiz. {\bf 54}, 1568 (1968)
[Sov.Phys.JETP {\bf 27}, 839 (1968)].

\bibitem{SHALAEV}
B.N.Shalaev, Fiz. Tverd. Tela (Leningrad) {\bf 32} 3586 (December 1990),
[Sov. Phys. Solid State {\bf 32} (12) 2079 (1990)].
\end{thebibliography}
\end{document}